\documentclass[aps,prb,twocolumn,groupedaddress,draft,intlimits,amsmath,amssymb,floats]{revtex4}

\usepackage{bm}
\usepackage[final]{graphicx}
\usepackage{epsfig}
\usepackage{subfigure}

\begin{document}

\title{Numerically exploring the 1D-2D dimensional crossover on spin dynamics in the doped Hubbard model} 

\author{Y. F. Kung$^{1,2}$}
\author{C. Bazin$^{2,3}$}
\author{K. Wohlfeld$^{2,4}$}
\author{Yao Wang$^{2,5}$}
\author{C.-C. Chen$^6$}
\author{C.J. Jia$^2$}
\author{S. Johnston$^{7,8}$}
\author{B. Moritz$^{2,9}$}
\author{F. Mila$^3$}
\author{T. P. Devereaux$^{2,10}$}

\affiliation{$^1$Department of Physics, Stanford University, Stanford, CA 94305, USA}
\affiliation{$^2$Stanford Institute for Materials and Energy Sciences, SLAC National Accelerator Laboratory and Stanford University, Menlo Park, California 94025, USA}
\affiliation{$^3$Institute of Physics, Ecole Polytechnique Federale de Lausanne (EPFL), CH-1015 Lausanne, Switzerland}
\affiliation{$^4$Institute of Theoretical Physics, Faculty of Physics, University of Warsaw, Pasteura 5, PL-02093 Warsaw, Poland}
\affiliation{$^5$Department of Applied Physics, Stanford University, California 94305,  USA}
\affiliation{$^6$Department of Physics, University of Alabama at Birmingham, Birmingham, Alabama 35294, USA}
\affiliation{$^7$Department of Physics and Astronomy, University of Tennessee, Knoxville, TN 37996, USA}
\affiliation{$^8$Joint Institute for Advanced Materials, The University of Tennessee, Knoxville, TN 37996, USA}
\affiliation{$^9$Department of Physics and Astrophysics, University of North Dakota, Grand Forks, ND 58202, USA}
\affiliation{$^{10}$Geballe Laboratory for Advanced Materials, Stanford University, Stanford, CA 94305, USA}

\begin{abstract}
Using determinant quantum Monte Carlo (DQMC) simulations, we systematically study the doping dependence of the crossover from one to two dimensions and its impact on the magnetic properties of the Hubbard model.  
A square lattice of chains is used, in which the dimensionality can be tuned by varying the interchain coupling $t_\perp$.  
The dynamical spin structure factor and static quantities, such as the static spin susceptibility and nearest-neighbor spin correlation function, are characterized in the one- and two-dimensional limits as a benchmark.  
When the dimensionality is tuned between these limits, the magnetic properties, while evolving smoothly from one to two dimensions, drastically change regardless of the doping level.
This suggests that the spin excitations in the two-dimensional Hubbard model, even in the heavily doped case, cannot be explained using the spinon picture known from one dimension.  
The DQMC calculations are complemented by cluster perturbation theory studies to form a more complete picture of how the crossover occurs as a function of doping and how doped holes impact magnetic order.
\end{abstract}


\maketitle

\section{Introduction}\label{sec:intro}
A common feature of many strongly correlated electron systems, such as the cuprate high-temperature superconductors, is layers of planes where most of the interesting physics occurs.  
This has motivated extensive studies of two-dimensional models.  
However, quasi-one-dimensional materials whose internal crystalline structure is known to be made of weakly coupled chains, such as $\text{SrCuO}_2$, KCuF$_3$, and the organic Bechgaard salts, also exist and provide an alternative perspective on properties such as magnetism.\cite{book1kimsov,srcuo2,expx,expx2}  
Indeed, the dimensionality of the system under consideration plays a crucial role in its behavior.  
At the microscopic level, dimensionality impacts the role of interactions.  
In two dimensions, electrons have a much larger number of paths to avoid one another than in one dimension, where they have to interact.  
This difference drastically modifies the physics, as single-particle excitations can be described in terms of Landau quasiparticles in two dimensions but not in one.  
For example, on a chain, only collective spin and charge excitations are possible, leading to spin-charge separation that has been observed experimentally\cite{expespincharge} and has important consequences for the magnetic properties.  
Hence elucidating how the system changes as a function of dimensionality can provide a deeper understanding of the properties themselves.

The Hubbard model provides a simple, unified framework that describes one-, two-, and quasi-one-dimensional correlated electron systems, incorporating the effects of electron hopping and Coulomb interactions.  
It can be solved analytically in one dimension, with one of two approaches.  
Because the low-energy physics of the Hubbard chain belongs to the one-dimensional Luttinger liquid (LL) universality class (except for the charge sector at half-filling), the first approach uses the approximate bosonisation scheme to calculate the spectrum of the model and show some of its most prominent properties, such as the aforementioned spin-charge separation.\cite{giamarchi:book, 1d:lljv}  
Low-energy properties and asymptotic correlation functions can also be evaluated.  
The second approach uses the analytically exact Bethe ansatz,\cite{hongrois3} which provides a way not only to compute the spectrum of the system, but also to evaluate the LL parameters.  
Together, these two methods demonstrate that spin-charge separation and its consequences for physical observables are the signature of one-dimensional physics.  
Yet, although the Hubbard chain can be solved exactly, obtaining information about static and dynamical correlations still relies strongly on numerical methods,\cite{giamarchi:finiteT} and the spin dynamics of the arbitrarily doped system has not been explored in detail.\cite{Ha1994, assaadyn1d2, 24spinon}

While the one-dimensional Hubbard model can be solved analytically, the two-dimensional case lacks exact solutions and is not yet fully understood due to the complexity of the physics it describes, necessitating the use of numerical techniques such as quantum Monte Carlo (QMC),\cite{BSS_PRD_1981,hirsh,vieuxqmcassad,mc2d1,qmcupd} exact diagonalization,\cite{Dagotto_RMP_1994,ARPACK} cluster perturbation theory,\cite{Senechal_PRL_2000,Senechal_PRB_2002} and dynamical mean-field theory.\cite{Georges_RMP_1996}  
Computational studies have established that the half-filled system is a $(\pi/a,\pi/a)$ antiferromagnetic (AF) insulator\cite{komski} and that the ground state is a Mott insulator in the strong coupling limit.  By the Mermin-Wagner theorem, long-range order (LRO) cannot exist at finite temperatures, implying that a gap opens and correlation functions are exponentially damped at $(\pi/a,\pi/a)$.  However, strong AF correlations are still present and influence physical observables even at finite temperatures.  Indeed, at $(\pi/a,\pi/a)$, the static correlation functions exhibit a dominant mode and the spin excitation spectrum has strong intensity.\cite{hirsh,cj}  Hence, for simplicity, we will use the term ``LRO" when referring to cases in which the AF correlations extend across the finite-size cluster.  Probing the spin dynamics at half-filling corresponds to the creation of one magnon, so the physics is approximated well by linear spin-wave theory.  A good understanding of the half-filled strongly coupled case is possible because spins remain localized. 

When holes are doped into the system, though, the interplay between AF order and hole delocalization complicates the situation.  Upon doping, AF LRO rapidly disappears \cite{dyntj} and the doped system exhibits a large variety of phases that compete or cooperate with one another.
As a consequence, understanding the spin dynamics of the doped two-dimensional Hubbard model remains a nontrivial task. 
In fact, it was only due to recent resonant inelastic x-ray experiments on doped cuprates, which
suggested (rather surprisingly) that collective spin excitations may persist up to high doping level in some regions of the Brillouin zone,\cite{LeTacon_NatPhys_2011,LeTacon_PRB_2013,Dean_PRL_2013,Dean_NatMat_2013,Lee_NatPhys_2014} that this problem was studied in greater detail.\cite{cj,Kung_PRB_2015}

Because the Hubbard model exhibits markedly different behavior in one and two dimensions, examining how the system crosses over between them provides insights into properties such as magnetism.  The crossover can be modeled with a system of coupled chains, where an interchain coupling, $t_\perp$, tunes a transition from effectively decoupled chains with confined electrons ($t_\perp=0$) to a deconfinement of electrons throughout the lattice ($t_\perp=t$ for a two-dimensional system).  
Studies of quasi-one-dimensional systems use a variety of analytical and numerical methods to calculate single-particle and two-particle processes.  The analytical approaches generally rely on renormalization group procedures and methods similar to the field theories used in one dimension,\cite{1d:lljv} while the numerical methods include DMFT, QMC, and variational cluster approximation (VCA).\cite{xover:assaad,giamarchix1,Raczkowski_PRB_2015,Lenz_PRL_2016}

The interesting question is at which point the transition from one- to two-dimensional character occurs.  Because the one-dimensional system is a Mott insulator with a gapped charge sector at half-filling but turns into a LL when doped, the two cases must be studied separately.  In the half-filled system with an intermediate $U$ value ($\sim 3t$), tuning $t_{\perp}$ can trigger a phase transition. 
A DMFT study has shown that for sufficiently small $U$ and $t_{\perp}$, a Fermi surface forms as $t_{\perp}$ increases.\cite{giamarchix1}  In addition, sufficiently strong next-nearest-neighbor hopping can prevent spin-density waves from opening a gap in two dimensions when $U$ has an  intermediate value.\cite{mitx}  Thus, the frustrated two-dimensional system is gapless at half-filling, and a metal-insulator transition occurs as $t_{\perp}$ is increased, as shown by QMC.\cite{assaadx2}
VCA and cluster DMFT have also studied the impact of the dimensional crossover on Mott quantum criticality.\cite{Lenz_PRL_2016}

When the system is doped, it has gapless charge sectors in both one and two dimensions.  Renormalization group and perturbative approaches show that interchain single-particle motion is controlled by the parameter $\alpha=(K_\rho+1/K_\rho-2)/4$, where $K_\rho$ is the LL parameter.\cite{bourbonx2,mila1,fabrizio}  When $\alpha \le 1$, $t_{\perp}>0$ is sufficient for interchain coherent motion.\cite{schulzx1}  When $\alpha>1$, the particles remain confined to the chains.  Analytical results suggest that the Hubbard model with finite $U$ should have $\alpha<1/8$ and hence interchain motion for finite $t_\perp$.\cite{schulzx1}  
However, numerical studies demonstrate that the situation is more complicated.  A QMC study finds that electrons can be confined for intermediate values of $\alpha$ smaller than 1.\cite{mila2}  Another study shows that as $t_{\perp}$ increases, coherent interchain motion develops, and the spectral function evolves from a LL form with decoupled chains and spin-charge separation towards a Fermi-liquid-like one with two-dimensional character and well-defined quasiparticle peaks.\cite{akwx}  However, for intermediate values of $t_\perp$, LL features remain present at high energies. 
Similarly, a DMFT study demonstrates that with an intermediate $U$ value ($U=4t$), lowering the temperature can trigger a transition from a LL to a Fermi liquid for $t_{\perp}>0$.\cite{giamarchix1}  These numerical results show that LL features can be observed even for finite interchain coupling.

Thus far, studies of magnetism in the dimensional crossover regime have focused on the half-filled system. 
As discussed already, the half-filled two-dimensional, but not one-dimensional, system is an AF insulator that develops LRO at zero temperature.  A question of great interest is the value of $t_{\perp}$ necessary to induce this LRO in the chain.\cite{1d:lljv}  As yet there is no definitive answer, but a renormalization group study\cite{lrox1} and multiple numerical studies\cite{kimx1,nolro1} of the anistropic Heisenberg model, as well as a QMC study of the intermediate-$U$ Hubbard model,\cite{xover:assaad} have suggested that any $t_{\perp}>0$ is sufficient to recover LRO.
Recently, determinant quantum Monte Carlo (DQMC) has been used to explore the evolution of spin and charge dynamics in the Hubbard model.\cite{Raczkowski_PRB_2015}  However, despite these studies, the interplay between doping and magnetism has not yet been examined in the framework of the dimensional crossover.

The goal of this study is to shed light on the magnetism of the strongly correlated doped Hubbard model as it transitions from one to two dimensions.  In addition to providing a means of comparison between the doped magnetism and magnetic excitations in one and two dimensions, the effect of the dimensionality on the doped magnetism of the Hubbard model is interesting in itself.
Hence, the magnetic properties are examined on a lattice of coupled chains where the interchain coupling is varied, in order to build upon previous results in one and two dimensions, and to elucidate magnetic properties such as the spin dynamics that are not yet well understood.  As suggested in an earlier study,\cite{cj} short-range spin correlations can also provide insight into the effect of doping on both magnetic order and spin excitations.  These quantities are computed via DQMC\cite{hirsh,mc2d1,BSS_PRD_1981,raimundo} and the maximum entropy method (MaxEnt) of analytic continuation.\cite{maxent}

The paper is organized as follows.  In Section II, the Hubbard model is presented together with the numerical methods used to carry out the simulations. Section III discusses the static and dynamic spin properties in one and two dimensions as a benchmark before Section IV explores the doping dependence of the dimensional crossover. Finally, Section V summarizes the main outcomes of this study and discusses further perspectives.

\section{Model and Methods}
The single-band Hubbard Hamiltonian\cite{Anderson_PR_1959,Hubbard_PRSLA_1963,Dagotto_RMP_1994} describes strongly correlated electrons on a lattice:
\begin{eqnarray}
H&=&\sum_{\langle i,j \rangle \sigma} t_{ij}(c^\dagger_{i\sigma} c^{\phantom{\dagger}}_{j\sigma} + h.c.)-\mu\sum_{i\sigma}n_{i\sigma}\nonumber\\
&&+U\sum_i \bigg(n_{i\uparrow}-\frac{1}{2}\bigg)\bigg(n_{i\downarrow}-\frac{1}{2}\bigg),
\end{eqnarray}
where $c^\dagger_{i\sigma}$ ($c^{\phantom{\dagger}}_{i\sigma}$) creates (annihilates) a particle with spin $\sigma$ on site $i$, and $n^{\phantom{\dagger}}_{i\sigma}=c^\dagger_{i\sigma}c^{\phantom{\dagger}}_{i\sigma}$ is the number operator.  The nearest-neighbor hoppings along the same chain and between chains are controlled by $ t_{ij} \equiv t$ and $ t_{ij} \equiv t_\perp$, respectively.  
The longer range hoppings are all set to zero except in the two-dimensional system, for which we may also consider the case with a finite next-nearest neighbor hopping $t_{ij} \equiv t'$.  $U$ is the on-site Coulomb interaction that penalizes double occupancy, and $a=1$ is the unit of length.  We work with $U=8t$, so the ground state is a strongly correlated Mott insulator in the undoped system, and measure energies in units of $t$.\cite{Lee_RMP_2006}  The chemical potential $\mu$ is adjusted to give the desired doping.  The model exhibits particle-hole symmetry, and the hole doping level can be defined as $p=1-n$, where $n$ is the electron density.  

In this study, properties of the Hubbard model are calculated using DQMC,\cite{hirsh,mc2d1,BSS_PRD_1981,raimundo} a numerically exact, auxiliary-field technique that computes observables from imaginary-time Green's functions as  

\begin{eqnarray}
\langle \hat{O} \rangle=\frac{\mathrm{tr} [\hat{O}e^{-\beta H}]}{\mathrm{tr} [e^{-\beta H}]},
\end{eqnarray}
with the imaginary-time interval $[0,\beta]$ divided into $L$ slices of width $\Delta \tau$.  The Hamiltonian can be rewritten in terms of the non-interacting and interacting pieces, and the exponential decomposed using the Trotter approximation

\begin{eqnarray}
e^{-L \Delta\tau H} \approx (e^{-\Delta\tau K}e^{-\Delta\tau V})^L,
\end{eqnarray}
where $K$ contains quadratic terms and $V$ is the quartic interaction term.  Terms in the expansion of order $O(\Delta\tau^2)$ and higher are dropped.  In this study, a sufficiently small time slice was used such that no significant $\Delta\tau$ errors were found.

A Hubbard-Stratonovich (HS) transformation 
\begin{eqnarray}
&&e^{-\Delta\tau U (n_{i\uparrow}-\frac{1}{2})(n_{i\downarrow}-\frac{1}{2})}\nonumber\\
&&=\frac{1}{2}e^{U \Delta\tau/4}\sum\limits_{s_{i,l}=\pm1} s_{i,l} e^{-\Delta\tau \lambda s_{i,l} (n_{i\uparrow} - n_{i\downarrow})},
\end{eqnarray}
is used to rewrite $V$ in quadratic form, at the cost of introducing a new HS field $s_{i,l}=\pm1$ at each lattice site $i$ and time slice $l$.  The relation $\cosh{(\Delta\tau \lambda)}=\exp{(\Delta\tau U/2)}$ defines $\lambda$.  The partition function can now be calculated as
\begin{eqnarray}
Z = \sum\limits_{s_{i,l} =\pm1} \det{M^{+}} \det{M^{-}},
\end{eqnarray}
where

\begin{eqnarray}
M^{\sigma} &=& I + B^\sigma_LB^\sigma_{L-1}...B^\sigma_1
\end{eqnarray}
and
\begin{eqnarray}
B^\pm_l = e^{\mp\Delta\tau \lambda v(l)} e^{-\Delta\tau K},
\end{eqnarray}
and $v(l)$ is a diagonal matrix with $s_{i,l}$ as the $i^{th}$ element.
The Monte Carlo sampling is performed over the HS field configurations, each of which has a weight of $P(s)=\det{M^{+}} \det{M^{-}}/Z$.  This is used to compute the Green's function, which is in turn used to compute all other quantities via Wick's theorem.  Since all observables are calculated in terms of imaginary time, they must be analytically continued to real frequencies for comparison to experiments.  In this study we employ MaxEnt, which uses Bayesian statistical inference to determine the most probable spectral density given an imaginary-time correlator.\cite{maxent} 

\begin{figure}[t!]
	\includegraphics[scale=0.6]{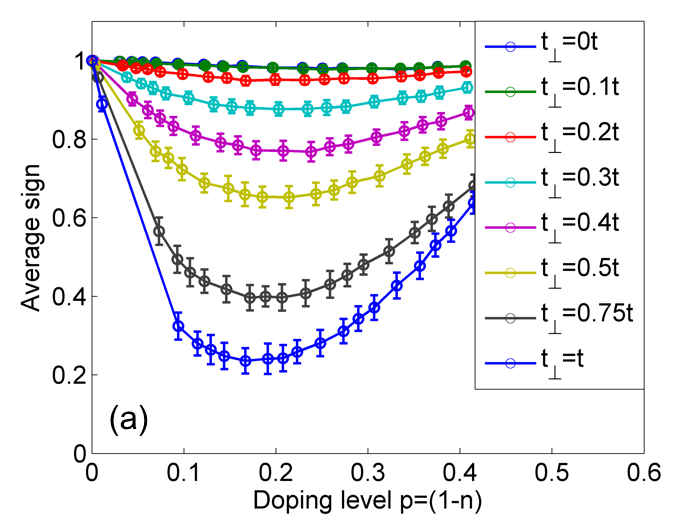}
	\caption{The doping dependence of the average fermion sign is shown for different values of interchain coupling at $\beta=3/t$.}
	\label{fig:sign_2d}
\end{figure}

DQMC has the advantages of being numerically exact and of accessing relatively large system sizes, but in general it suffers from a fermion sign problem.\cite{mc2d1,Iglovikov_PRB_2015,Iazzi_PRB_2016}  Because the algorithm does not track the order of the operators, a negative sign from the fermion anticommutation relations remains undetermined and all observables must be divided by the average fermion sign $\langle f_\mathrm{sgn}\rangle$ as 
\begin{eqnarray}
\langle \hat{O} \rangle &=& \frac{\sum\limits_{s_{m,l}} \hat{O} f_\mathrm{sgn} P(s)}{\sum\limits_{s_{m,l}} f_\mathrm{sgn} P(s)}=\frac{\langle O f_\mathrm{sgn}\rangle}{\langle f_\mathrm{sgn}\rangle},\nonumber\\
f_{sgn} &=& \mathrm{sign}(\det{M^{+}} \det{M^{-}}).
\end{eqnarray}
Statistical fluctuations become more significant as the average sign decreases; hence its value controls accessible parameter regimes.  

Figure~\ref{fig:sign_2d} systematically explores the average fermion sign for different doping levels and interchain hoppings.  At half-filling, particle-hole symmetry protects the sign such that it is always 1.  Away from half-filling, however, the average sign is suppressed exponentially as the temperature decreases and the system size increases.\cite{Iglovikov_PRB_2015}  Despite doping, the average sign remains close to 1 in one dimension due to the small number of available hopping pathways (and opportunities for ambiguities in sign).  As the interchain coupling $t_\perp$ and dimensionality increase, the system smoothly evolves from one to two dimensions, where the average sign is strongly reduced upon doping.  Doped two-dimensional systems require higher temperatures to have a non-vanishing average sign, constraining our simulations to $\beta=3/t$ in general.

\section{Spin Properties in One and Two Dimensions}\label{sec:1d}
\subsection{One-Dimensional Case}
We begin with a review of the one-dimensional case, systematically studying its static and dynamic magnetic properties for comparison to the two-dimensional case.  The static spin properties elucidate the effects of doping on magnetic order and short-range correlations.  The peak position of the static spin susceptibility $\chi(q)$ shifts with doping as $2k_F=n\pi$.  The peak intensity is highest at half-filling, $q=\pi$, suggesting strong AF correlations, and decreases upon doping, indicating that the $2k_F$ spin-density wave weakens as its mode changes.\cite{1d:qmc1}  
This weakening is confirmed by a decrease in the magnitude of the spatial spin-spin correlation function $\langle\textbf{S}_i\cdot\textbf{S}_j\rangle$ upon doping.  Finite temperature destroys the quasi-LRO, implying an exponential decay of correlations with a finite correlation length $\xi$, which can be extracted from $\langle\textbf{S}_i\cdot\textbf{S}_j\rangle$.  At half-filling, $\xi$ is in good agreement with the Bethe ansatz.\cite{corrba}  The small correlation length suggests that short-range correlations play an important role in the observed magnetic properties.  

Short-range correlations can be examined via the nearest-neighbor (NN) spin correlation $\langle\textbf{S}_0\textbf{S}_1\rangle$, which shows that low levels of hole doping reduce the local density of spins without additionally destroying magnetic order, as is consistent with the importance of short-range correlations upon doping.
However, for higher doping levels, the NN correlation decreases more rapidly than expected from a local static picture.
Naively, one would think that once most electrons have been removed, hole delocalization should have a reduced impact on the magnetic order. However, spin-charge separation complicates the situation, as doping a hole corresponds to adding both a holon and a spinon.  They can be thought of as acting independently (due to spin-charge separation); thus the holon would delocalize without affecting (and destroying) the magnetic order, explaining the behavior of the NN spin correlations up to intermediate doping levels.  However, spinons may act like domain walls and cause additional destruction of magnetic order.  This picture would be consistent with the faster decay of the NN spin correlations at large doping. The transition means that the way in which doped holes destroy magnetic order evolves as electron density decreases.

\begin{figure}[b!]
	\includegraphics[width=0.8\columnwidth]{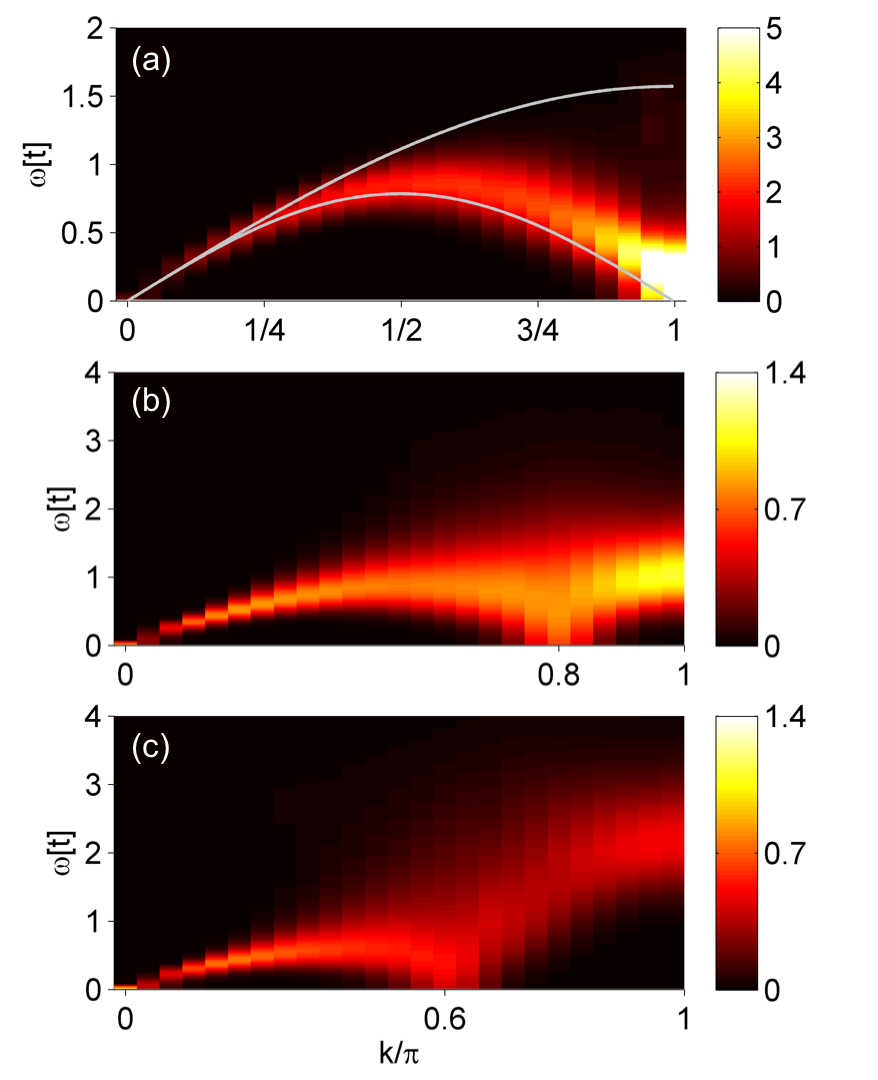}
	\caption{False-color plots of the dynamical spin structure factor are shown at (a) $n=1.0$, (b) $n=0.8$, and (c) $n=0.6$.  The solid lines in panel (a) correspond to the upper and lower boundaries of the two-spinon continuum.  Calculations are performed on a 48-site Hubbard chain with $U=8t$ and $\beta = 15/t$ to access the expected low-temperature behavior.}
	\label{fig:1d_dyn}
\end{figure}

The dynamical spin structure factor, $S(\textbf{q},\omega)$, provides energy- and momentum-resolved information about magnetic properties.  Fig.~\ref{fig:1d_dyn}(a) shows that the half-filled system exhibits a continuum, confirming that a spin flip decays into independent spinons, as expected in the presence of spin-charge separation.\cite{srcuo22}  
The spectral intensity is well described by the two-spinon continuum computed from the Bethe ansatz for the Heisenberg model (solid lines), with spectral weight concentrated at the lower boundary due to suppression of itinerancy effects and dominant spin excitations at low energy due to virtual electron hopping for larger $U$.\cite{itinerant2spinons}  The spectrum is broadened by finite temperature, which also destroys quasi-LRO and opens a small gap at $\pi$.  Nevertheless, the high spectral intensity at $2k_F=\pi$ confirms the presence of strong AF correlations. 
In addition, despite finite-size and finite-temperature effects, the spin velocity arising from the LL formalism agrees well with the Bethe ansatz prediction,\cite{giamarchi:book} suggesting that at half-filling, the spin dynamics in the strong coupling limit is described well by the Bethe ansatz solution of the Heisenberg chain at zero temperature.

In the doped system [Figs.~\ref{fig:1d_dyn}(b) and (c)], spectral intensity both decreases and broadens, while the spectrum hardens overall.\cite{assaadyn1d2}  In addition, the damping of intensity is the largest at $\pi$.\cite{24spinon}  The continuum indicating the presence of spin-charge separation still can be seen, suggesting that the doped chain also exhibits characteristics of a LL, which is further confirmed by the soft mode at $2k_F=n\pi$ (where $n$ is the electron density). This shift in $2k_F$ away from $\pi$ with doping also explains why the spectrum becomes increasingly gapped at $k=\pi$: It no longer corresponds to the soft mode of the spin-density wave of the doped system.  As in the half-filled case, the spin velocities can be extracted and are found to be in good agreement with theoretical values.\cite{giamarchi:book} 

The qualitative behavior of the doped spectra is very different from that of the two-spinon continuum, which is symmetric about $2k_F$. As four-spinon processes are expected to contribute more as doping increases,\cite{24spinon} this qualitative observation seems to confirm that processes other than two-spinon ones are involved in the spin dynamics and are enhanced by doping. Moreover, as opposed to the half-filled case, they do not seem to overlap simply with the two-spinon continuum. 

In order to explain the overall hardening of the spin excitation spectrum upon doping, naively one could study it in terms of the local static picture. Although there is no LRO in one dimension, the robustness and the relatively strong intensity of the continuum at high doping levels are likely related to strong short-range correlations.\cite{24spinon}  

\subsection{Two-Dimensional Case}
Extensive DQMC studies have already been performed to characterize static and dynamic magnetic properties in two dimensions, which we review briefly.  Both the static spin susceptibility and real-space spin-spin correlation function provide evidence for the presence of AF order at half-filling and its destruction with doping, as expected.\cite{hirsh}

The NN spin correlation $\langle \textbf{S}_0\cdot\textbf{S}_1\rangle$ explores the interaction of doped carriers with the magnetic background.  
When $t^\prime=0$, a discrepancy in $\langle \textbf{S}_0\cdot\textbf{S}_1\rangle$ from what would be expected in the simple local static picture is observed up to intermediate hole-doping levels. 
In fact, this result is not surprising, since in the spin polaron picture~\cite{Martinez1991}, introducing holes into a magnetically ordered background causes a strong reduction of magnetic correlations that is larger than the simple reduction in magnetic moments.\cite{Bonca1989, Winterfeldt1998,Vladimirov2009, Kar2011} 
However, once the number of doped holes exceeds $50\%$, the density of electrons is small enough for the holes to delocalize without breaking AF bonds.
Then hole delocalization no longer competes with magnetic correlations and the NN spin correlation corresponds to the local static picture.

A distinct situation occurs once the next-nearest-neighbor (NNN) hopping $t^\prime = -0.3t$ is switched on (note that such a value of the NNN hopping is typically chosen so that the Hubbard
model constitutes a more realistic description of the cuprates).\cite{cj} On the hole-doped side, the effect of $t^\prime$ is to favor sublattice mixing, which leads to an enhanced destruction of AF order.  On the electron-doped side, finite $t^\prime$ supports AF correlations,\cite{polaron} so that the local static picture can explain the hardening of the spin excitation spectrum. The latter, rather surprising, result strongly depends on the choice of $t^\prime$:
It can only be obtained for strong enough $t^\prime$, i.e. a negative but much smaller value $|t^\prime| \ll 0.3 t$ would not be enough. 

\begin{figure}[t!]
	\includegraphics[scale=0.7]{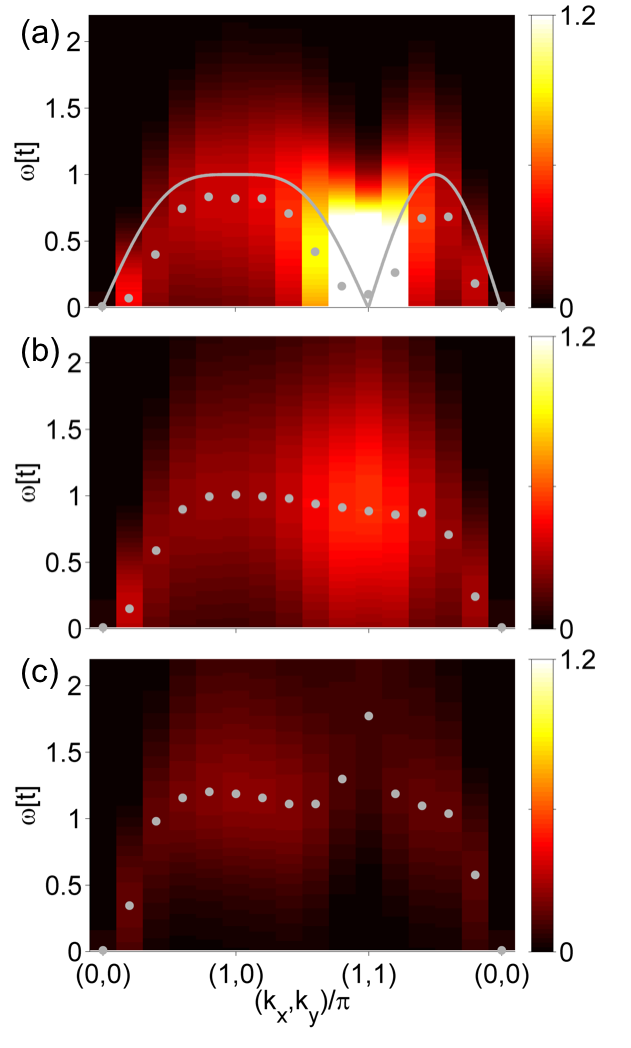}
	\caption{False-color plots of the dynamical spin structure factor $S(\textbf{q},\omega)$ along high-symmetry directions are shown at (a) $n=1.0$, (b) $n=0.8$, and (c) $n=0.6$ electron density on a $10\times 10$ cluster with $U=8t$ and $\beta=3/t$.  The spectra at $n=0.9$ and $n=0.7$ have also been obtained and interpolate smoothly between the spectra shown here. The dots correspond to the maximum intensity for a given momentum point, and the solid line at half-filling is the linear spin-wave dispersion calculated for a nearest-neighbor Heisenberg model with $J=4t^2/U=0.5t$.}
	\label{fig:2d_dyn}
\end{figure}

Figure~\ref{fig:2d_dyn} shows $S(\textbf{q},\omega)$ at different doping levels, with dots indicating the maxima of spectral intensity. 
Despite thermal broadening and renormalizations from quantum effects, in general the spin excitation spectrum exhibits the features expected at half-filling [Fig.~\ref{fig:2d_dyn}(a)]. The intensity maxima reproduce the linear spin-wave dispersion (calculated for a nearest-neighbor Heisenberg model with $J=4t^2/U=0.5t$ for $U=8t$) up to a multiplicative factor.  However, $(\pi,\pi)$ has the strongest intensity, indicating that AF order  builds up at half-filling despite the absence of true LRO in the numerical simulation.

As the doping level increases [Figs.~\ref{fig:2d_dyn}(b) and (c)], the spectrum hardens and spectral intensity decreases. These effects are most pronounced at $(\pi,\pi)$, which is affected by the destruction of AF order.
Along the direction $(0,0)\rightarrow (\pi,0)$, spectral intensity remains significant even at $40\%$ doping, compared to that at other momentum points, similarly to the finite $t'$ case reported in Ref.~\onlinecite{cj}.
Moreover, the spectrum hardens along the $(0,0)\rightarrow (\pi,0)$ direction. This is a counterintuitive result, since an overall softening has been suggested in the literature.\cite{Vladimirov2009, Kar2011, Chen2013}
However, a study comparing spin susceptibilities calculated by DQMC and the random phase approximation, which was developed for weakly interacting systems, demonstrated that this discrepancy may be caused by significant correlations that persist to higher doping levels.\cite{Kung_PRB_2015}

\section{Magnetic Properties in the Dimensional Crossover}\label{sec:crossover}
Depending on the dimensionality, the magnetic response of the Hubbard system can be very different -- e.g.
the spin dynamical structure factor shows the onset of spinon continua 
in one dimension, whereas surprisingly stable $S=1$ collective spin excitations (magnons) are observed in two dimensions, even at high doping levels.  Here we present results for the crossover from one to two dimensions and explore the doping dependence of magnetic properties as a function of dimensionality.  Unless otherwise noted, calculations are performed on a $10\times 10$ cluster with $U=8t$, $\beta = 3/t$, and interchain coupling values $t_{\perp}/t=0.0,0.1,0.2,0.4,0.6,0.8,1.0$ (of which representative values are shown).

We note that the lattice consists of decoupled chains when $t_{\perp}=0$.  Thus, physical properties only have spatial dependence along the chain and are independent of the transverse direction.  In reciprocal space, the direction $(0,0)\rightarrow(\pi,0)$ effectively corresponds to a single chain.  Due to the transverse-direction independence, any properties along $(0,k)\rightarrow(\pi,k)$ will be identical to the those along the chain, and properties along $(k,0)\rightarrow(k,\pi)$ will be identical to those at $(k,0)$.  These considerations should be kept in mind when interpreting the dimensionality-dependent results. 

\subsection{Static Properties}\label{sec:crossover_static}
First we study how the static spin properties evolve upon tuning $t_{\perp}$.  As before, the static spin susceptibility provides an energy-integrated perspective, while the NN spin correlation function shows how doped holes interact with the magnetic background.

\subsubsection{Static Spin Susceptibility}
\begin{figure}[t!]
	\includegraphics[width=\columnwidth]{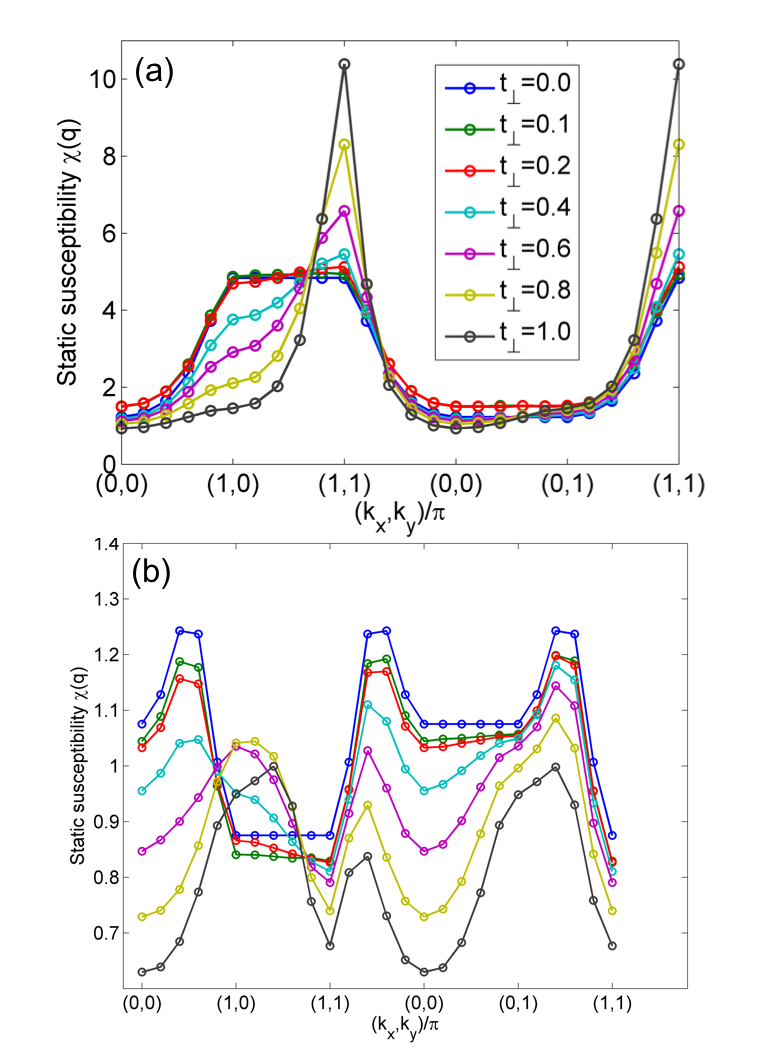}
	\caption{Static spin susceptibility $\chi(\textbf{q})$ for different values of $t_{\perp}$ at (a) half-filling and (b) $40\%$ hole doping. The magnetic order is transferred from one to two dimensions. The crossover only occurs for $t_{\perp}\geq 0.4t$ because of the high temperature.}
	\label{fig:sucx}
\end{figure}

 Figure~\ref{fig:sucx} shows the static spin susceptibility at half-filling and $40\%$ hole doping for different values of $t_{\perp}$.  For $t_{\perp}< 0.4t$, the susceptibility only has momentum dependence along the chains and remains almost flat along the transverse direction. This again indicates that the magnetic order mostly retains one-dimensional character for small $t_{\perp}$. 
 
Figure~\ref{fig:sucx}(a) shows that at half-filling, magnetic order is transferred smoothly from $(\pi,0)$, which corresponds to the spin-wave mode along the chain, to $(\pi,\pi)$ as interchain coupling increases.  As soon as $t_\perp > 0$, the static susceptibility is larger at $(\pi,\pi)$ than at $(\pi,0)$.  At large $t_{\perp}$, the intensity at $(\pi,\pi)$ is strongest, showing that AF order dominates and suggesting that increasing $t_{\perp}$ leads to spinon confinement and AF LRO.  Therefore, as has been deduced from the transfer of the spectral intensity in the spin dynamics, the system smoothly develops AF order as the chains are increasingly coupled.

In comparison with Fig.~\ref{fig:sucx}(a), the AF order at $40\%$ doping [Fig.~\ref{fig:sucx}(b)] has been reduced strongly.  In addition, the susceptibility at wave vectors that dominate in both one and two dimensions at half-filling are weakened significantly by doping. 
Along the $(0,0)\rightarrow (\pi,0)\rightarrow (\pi,\pi)$ direction, the soft mode in the spin dynamics expected in one dimension at $\textbf{q}=(2k_F,0)$ is not observed because of the high temperature.  Nevertheless, for $t_{\perp}\leq 0.2t$, the static spin susceptibility has a peak along the chain corresponding to the $2k_F$ one-dimensional mode. 
This observation clearly shows that for small enough $t_{\perp}$, the chains retain their one-dimensional character at $40\%$ doping. 

We note that this $2k_F$ mode is directly related to the soft mode expected in the spin dynamics. Thus, the static susceptibility provides us with a quantitative means of tracking the evolution from one- to two-dimensional character in the doped system.
Indeed, as $t_{\perp}$ increases above $0.4t$, the spectral intensity of the one-dimensional-like peak at $(2k_F,0)$ shifts smoothly toward $(\pi,\pi-q)$, which is expected in two dimensions for high doping. 

Therefore, studying the static spin susceptibility allows us to track more closely the evolution of the one-dimensional properties as $t_{\perp}$ is varied in the doped system.  At half-filling, low temperatures can be reached so that most physical observables can be used to track the dimensional crossover.  However, in the doped case, the sign problem restricts simulations to higher temperatures, making it difficult to observe one-dimensional behavior.  Thus the static susceptibility enables us to quantify the dimensional crossover. 

\subsubsection{Nearest-Neighbor Spin Correlations}
\begin{figure}[t!]
	\includegraphics[width=\columnwidth]{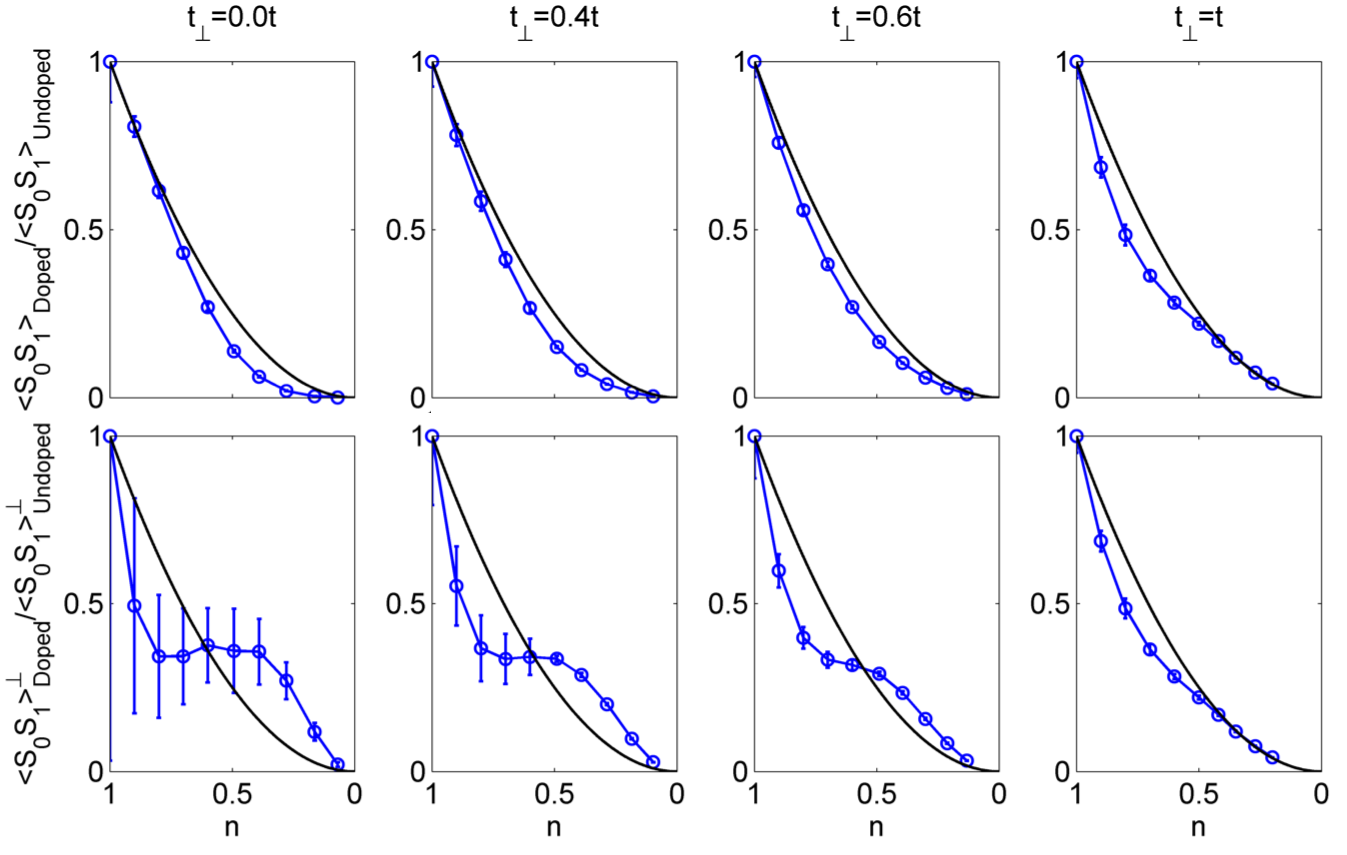}
	\caption{NN spin correlation as a function of electron density $n$ for different values of the interchain hopping $t_{\perp}$, shown both along (top row) and perpendicular to (bottom row) the chain.  Perpendicular to the chain, the spin correlation function for $t_{\perp}=0.1t$ is shown instead of that for $t_{\perp}=0t$, which is identically zero.  The values are very small close to one dimension, leading to larger error bars.  The solid lines show the $(1-p^2)$ curve predicted by the local static picture.}
	\label{fig:nnx}
\end{figure}

As already discussed, dimensionality has a strong impact on the way in which hole delocalization interacts with magnetic order as a function of doping.  In one dimension, hole delocalization destroys magnetic order by reducing spin density at low doping levels but causes a greater destruction at high doping levels.  In two dimensions, the opposite trend has been observed.  To examine how the behavior crosses over between these two limits, the longitudinal (along the chain) and transverse NN spin correlation functions are calculated as a function of doping for different values of $t_{\perp}$, as shown in Fig.~\ref{fig:nnx}.

\begin{figure*}[t!]
	\includegraphics[scale=0.7]{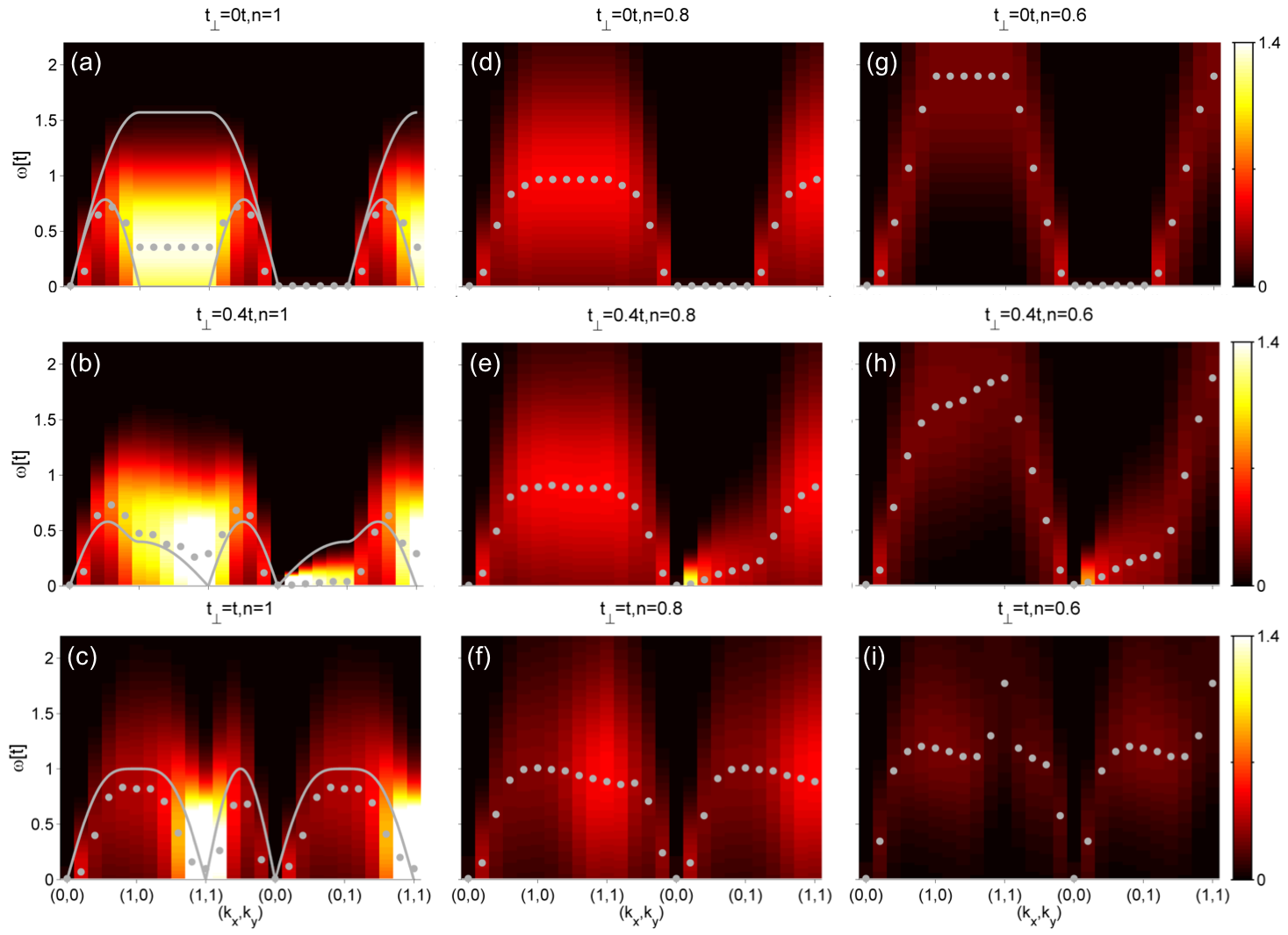}
	\caption{False-color plot of the dynamical spin structure factor $S(\textbf{q},\omega)$ along the main symmetry directions, for different doping levels and values of the interchain coupling, $t_{\perp}$.  At half-filling in one dimension, the solid line corresponds to the two-spinon continuum; for $t_{\perp}\neq 0$, it represents the linear spin-wave dispersion. The dots indicate the maximum intensity for each momentum point, which is given in units of $\pi$.}
	\label{fig:xdyn_doping}
\end{figure*}

Although the doping trend of the longitudinal correlations [Fig.~\ref{fig:nnx} (top row)] interpolates smoothly from one to two dimensions, the transverse  correlations [Fig.~\ref{fig:nnx} (bottom row)] show an unexpected trend.  After a rapid decrease to a plateau, upon increasing doping the transverse NN spin correlation function recovers, and its magnitude even exceeds the prediction from the local static picture.  This plateau suggests a transitory regime where doped holes induce processes that compensate the reduction of magnetic order caused by local spin density reduction.  It could be related to the spin excitations along the transverse direction: As shown in Fig.~\ref{fig:xdyn_doping}, the dispersion initially lifts up as the system is hole doped, but does not evolve significantly upon further doping (up to $40\%$).  

To check whether this doping dependence is a thermal effect, the same measurements could be performed at lower temperatures, decreasing $t_{\perp}$ to ameliorate the sign problem.  We note that it would be surprising for high temperatures to enhance correlations above the simple local spin density reduction curve.
Thus, the doping dependence of the NN spin correlation function is intriguing and highlights the importance of short-range correlations in understanding the properties of the doped system.

\subsection{Spin Dynamics: Half-Filling}
\begin{figure*}[t!]
	\includegraphics[scale=0.5]{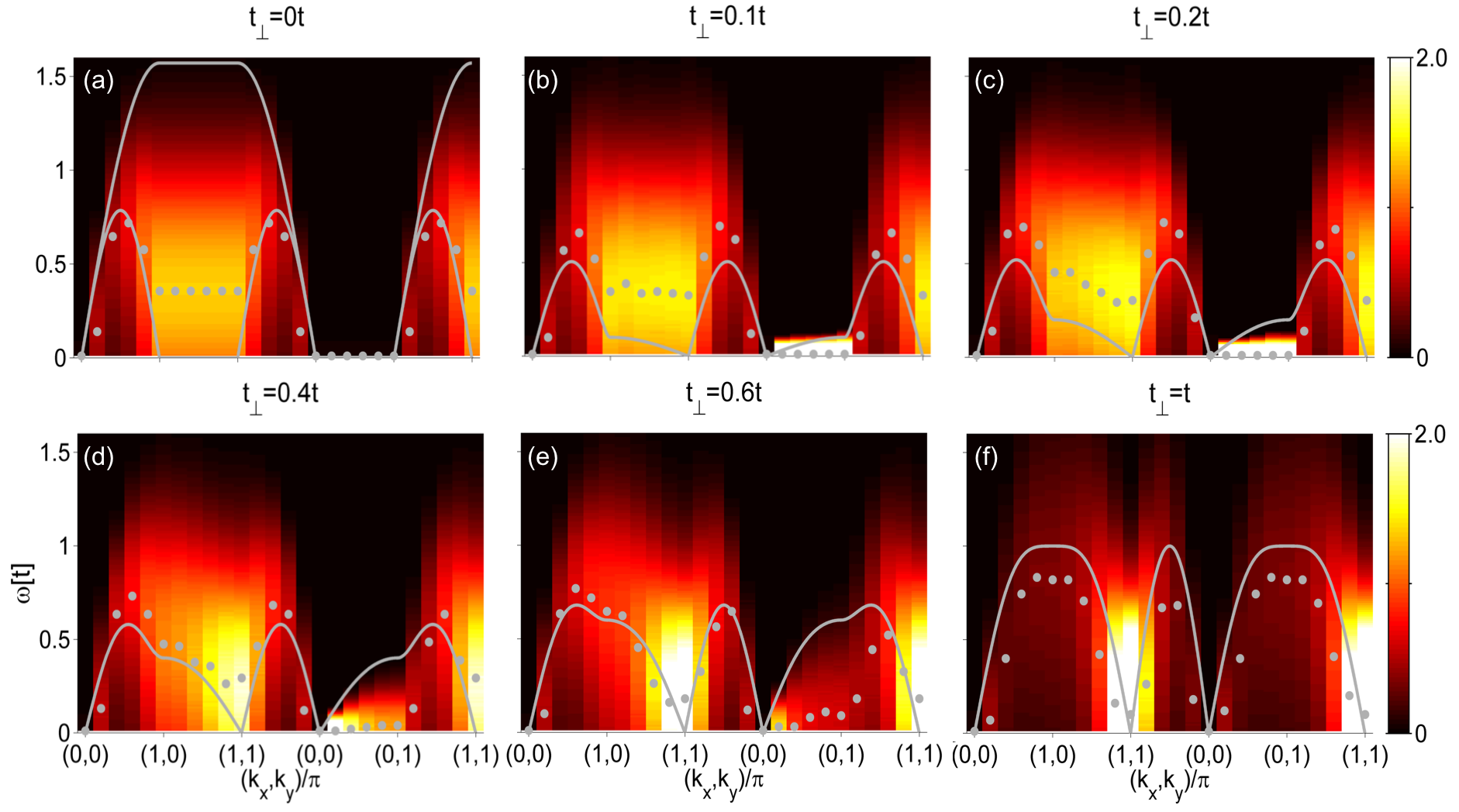}
	\caption{False-color plot of the dynamical spin structure factor, $S(\textbf{q},\omega)$, along the main symmetry directions at half-filling, for different values of $t_{\perp}$.  In one dimension, the solid line corresponds to the two-spinon continuum; for $t_{\perp}\neq 0$, it represents the linear spin-wave dispersion. The dots indicate the maximum intensity at each momentum point (in units of $\pi$).}
	\label{fig:xdynhf}
\end{figure*}

Figure~\ref{fig:xdynhf} shows the dynamical spin structure factor, $S(\textbf{q},\omega)$, for different values of $t_{\perp}$ at half-filling. The solid lines correspond to the boundaries of the two-spinon continuum for $t_{\perp}=0$, and to the linear spin-wave dispersion for $t_{\perp}>0$:
\begin{equation}
\omega(k_x,k_y)=\sqrt{\left(J+J_{\perp}\right)^2-\left(J\cos\left(k_x\right)+J_{\perp}\cos\left(k_y\right)\right)^2},
\end{equation}
where $J=4t^2/U$ and $J_{\perp}=4t_{\perp}^2/U$.  The spectra are broad due to the high simulation temperature, so dots indicating the intensity peaks provide a guide to the eye.

When $t_{\perp}=0$ [Fig.~\ref{fig:xdynhf}(a)], the system consists of $10$-site chains.  At $\beta=3/t$, the two-spinon continuum is barely distinguishable along $(0,0)\rightarrow (\pi,0)$, which corresponds to the dispersion along the chain.  However, the two-spinon continuum is clearly observed at $\beta=15/t$ (not shown), indicating that the $10$-site chain is long enough to capture one-dimensional spin dynamics at sufficiently low temperatures.

As $t_\perp$ increases [Figs.~\ref{fig:xdynhf}(b)-(f)], magnetic correlations transition from the ones known for the chain to those known for the two-dimensional lattice.  Spectral intensity shifts from $(\pi,0)$ toward $(\pi,\pi)$, suggesting the transition to AF LRO that is predicted at zero temperature for small transverse hopping.
In addition, peaks in the spectral intensity increasingly follow the linear spin-wave dispersion (up to a multiplicative factor), suggesting that coherent spin-wave excitations have replaced independent spinons, which have become confined. The only exception is along the direction $(0,0)\rightarrow (0,\pi)$, where the spectral intensity does not follow the linear spin-wave dispersion up to large values of $t_{\perp}$.\cite{xover:assaad}  Since low-energy spin excitations are present along this direction, strong thermal fluctuations probably impede the development of two-dimensional magnetism.
This transition from one- to two-dimensional magnetism was observed previously in Sec.~\ref{sec:crossover_static} in the static spin susceptibility.  

For small values of $t_{\perp}$, the spectrum mostly retains one-dimensional characteristics, such as the two-spinon continuum and weak momentum dependence along transverse directions.  This behavior is due to the simulation temperature, as high temperatures partially wash out the increase in dimensionality. 
Indeed, coherent spin waves form between $t_{\perp}=0.2t$ [Fig.~\ref{fig:xdynhf}(c)] and $t_{\perp}=0.4t$ [Fig.~\ref{fig:xdynhf}(d)] along certain directions.  In addition, the spectral intensity along the transverse direction $(0,0)\rightarrow (0,\pi)$ begins to disperse more strongly at $t_{\perp}=0.4t$. 

The effect of the dimensional crossover on the physical observables depends on the energy scale, as also reported in experiments.\cite{expx}  For example, the cuprates exhibit a spin-wave dispersion at low energies, and a spinon continuum at higher energies.\cite{expx2}  This transition temperature can be estimated naively as $T_{\mathrm{transition}}\sim t_{\perp}$;\cite{giamarchi:book} however, the transition affects single- and two-particle interchain processes differently\cite{bourbonx2} and is renormalized by interactions so that the effective crossover temperature is lower.\cite{giamarchi:book}
Despite the Coulomb interaction in this calculation, the naive non-interacting prediction of $T_{\mathrm{transition}}\sim t_{\perp}=t/3$ appears to hold at $\beta=3/t$.  The temperature dependence of the transition is confirmed by performing a simulation at $\beta=5/t$ for $t_{\perp}=0.2t$.  As expected, spectral intensity at $(\pi,\pi)$ is enhanced and the positions of the peak intensity  track the linear spin-wave dispersion more closely at $\beta=5/t$ than at $\beta=3/t$.  At the energy scales probed here, this crossover occurs smoothly.  These observations agree well with previous studies.\cite{xover:assaad}

\subsection{Spin Dynamics: Doping and Dimensionality}
In one dimension, the spin excitation spectrum hardens upon doping and develops a soft mode at $2k_F$.  In two dimensions, the spectrum exhibits an overall hardening but with a relatively persistent intensity along the axes in reciprocal space.  In order to study the effect of dimensionality on the doping dependence of the spin excitation spectra, Fig.~\ref{fig:xdyn_doping} shows the dynamical spin structure factor for different doping levels and interchain coupling strengths.

In one dimension, due to high temperature, the soft mode at $2k_F$ is not seen even though $20\%$ and $40\%$ hole doping correspond to $2k_F=n\pi=\frac{\pi}{5}$ and $\frac{2\pi}{5}$, respectively.  However, when the temperature is decreased to $\beta=15/t$ (not shown), the soft mode is observed, suggesting again that one-dimensional physics is present on a $10$-site doped chain even when obscured by high temperatures.  In fact, as discussed below, the $2k_F$ mode is seen in the static spin susceptibility even at $\beta=3/t$.

The temperature has the same effect on the crossover in the doped system as in the half-filled case, with the transition between one- and two-dimensional magnetism occurring between $t_{\perp}=0.2t$ and $t_{\perp}=0.4t$. Indeed, for $t_{\perp}<0.4t$, the energy scales of the spectra are almost unaffected and the momentum dependence along transverse directions of reciprocal space remains small. 
Generally, the dimensional crossover occurs smoothly in the spin dynamics, independent of the doping level, with the spectra transitioning gradually from the continuum in one dimension to showing more coherent spin excitations in two dimensions.  This suggests that the nature of the spin excitations gradually changes from spinon-like behavior at small $t_{\perp}$ to a magnon-like response at large $t_{\perp}$.  We will explore specific aspects of the crossover in greater detail in the following sections.

\subsubsection{Transverse Dispersion}  
Along the transverse direction, $(0,0)\rightarrow (0,\pi)$, finite $t_\perp$ causes the dispersion to lift up upon doping and the spectral intensity to spread more.  Doping the chains enhances the spin excitations and hence their two-dimensional character (they are dispersionless in one dimension but disperse in two dimensions). The effect of doping on the dimensionality dependence is clearly seen in Fig. \ref{fig:xdyn_doping}.  At half-filling, the spectral intensity peaks show a weak dispersion away from the linear spin-wave dispersion up to large values of $t_{\perp}$, but at high doping levels, spectral hardening is more sensitive to dimensionality, even for fairly small $t_{\perp}$.

A simple local picture offers insights into the enhanced dispersion upon doping.  Once holes are doped, the $t-J$ three-site term, in addition to the Heisenberg term, contributes to the spin dispersion.  In the transverse direction, these two terms depend differently upon $t_{\perp}$. For the Heisenberg exchange, the magnetic coupling involves a virtual hopping between nearest neighbors only, so $J_{\perp}=4t_{\perp}^2/U$ and is suppressed for small $t_{\perp}$.  On the other hand, there are three possible channels for the three-site term, which involves a virtual hopping between the nearest and next-nearest neighbors.\cite{cj}  In one of the channels, the coupling between spins depends linearly on $t_{\perp}$: $J^{\text{3-site}}_{\perp}=4tt_{\perp}/U$. Thus, at half-filling, the spin exchange scales as $t_{\perp}^2$ and is reduced, while in the doped case it scales as $t_{\perp}$. This linear dependence may explain the enhancement of the spin dispersion along $(0,0)\rightarrow (0,\pi)$.

The evolution of the peak intensity position of the dynamical spin structure factor at $(0,\pi)$ as a function of $t_{\perp}$ is shown in Fig.~\ref{fig:peaksize}(a) for different doping levels.  Near half-filling, the evolution is closer to $t_{\perp}^2$.  At high doping levels, the dependence becomes more linear even though it is obscured by variations from thermal broadening, which increases the variability of the peak position. Nevertheless, a clear transition occurs in the $t_{\perp}$ dependence between the half-filled and the doped cases. 

At half-filling, for small $t_{\perp}$, the energy scale of transverse spin excitations is too small compared to the temperature to observe a dispersion along $(0,0)\rightarrow (0,\pi)$.  However, once holes are doped, the contribution of the three-site term induces higher-energy excitations that are more easily probed at high temperatures, showing the importance of the term in understanding the physics of doped systems.  It also suggests that doping the system makes the crossover of the low-energy part of the spin excitation spectrum occur at a smaller interchain coupling value.  Thus, an anistropic lattice allows us to disentangle different processes and separate their contributions.

\subsubsection{Momentum-Dependent Crossover and Persistent Spin Excitations}  

\begin{figure}[t!]
	\includegraphics[scale=0.7]{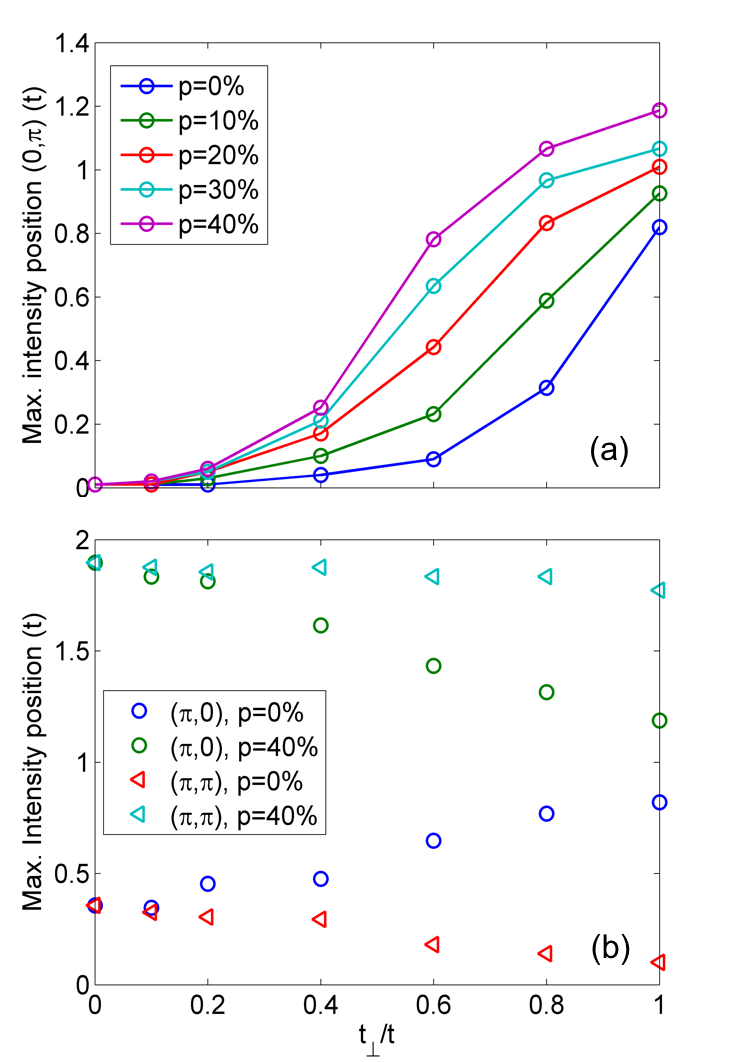}
	\caption{Energy of the spin excitations at different momenta as a function of $t_{\perp}$ for different doping levels.  Panel (a) focuses on the detailed doping dependence at $\textbf{k}=(0,\pi)$, while panel (b) compares the evolution of the energy of spin excitations at $\textbf{k}=(\pi,0)$ and $\textbf{k}=(\pi,\pi)$ at two different doping levels.}
	\label{fig:peaksize}
\end{figure}

The evolution of the spin dynamics at $(\pi,\pi)$ and $(\pi,0)$ can be compared at half-filling and at $40\%$ hole doping [Fig.~\ref{fig:peaksize}(b)].
In the half-filled case, once coherent spin excitations are recovered for $t_{\perp}\geq 0.4t$, their energy scales can be described with a linear spin-wave dispersion.  At $(\pi,\pi)$, AF order sets the energy scale, which has no significant $t_\perp$ dependence and remains nearly gapless.  At $(\pi,0)$, the spectrum hardens according to the linear spin-wave dispersion: $\omega(\pi,0)=2\sqrt{JJ_{\perp}}\sim t_{\perp}$. 

Although linear spin-wave theory can predict the half-filled behavior, it does not describe the $40\%$ hole-doped system.
As at half-filling, the energy scale of the $(\pi,\pi)$ spin excitations has only a small $t_{\perp}$ dependence.
In one dimension, $(\pi,\pi)$ is equivalent to $(\pi,0)$ and the hardening of the spectrum at $(\pi,\pi)$ thus corresponds to the hardening of the spectrum of the chain upon doping. It is remarkable that despite the very different natures of the spin excitations in one and two dimensions, the energy scale of this point remains the same as interchain coupling is increased.  On the other hand, the $(\pi,0)$ spin excitations soften as $t_{\perp}$ increases. Hence, as the spinons start to bind together, the energy cost of a spin flip along the chain direction decreases, and the energy scale of the spectrum smoothly interpolates from one to two dimensions.  Therefore, there is almost no crossover in the energy scale of the spin excitations at $(\pi,\pi)$ while one exists at $(\pi,0)$.

\subsubsection{Hardening of the Spectrum}
In two dimensions, the spectrum has been observed to harden upon doping, along the $(0,0)\rightarrow (\pi,0)$ direction.  In fact, as shown in Fig.~\ref{fig:xdyn_doping}, this behavior occurs for any given value of $t_{\perp}$.  Moreover, the energy scale of the hardening is always larger than that in two dimensions (ranging from $\sim 1.2t$ in two dimensions to $\sim 2t$ in one dimension). 
 Although the local static picture does not fully apply, it can still shed light on why the spectrum hardens more for $t_{\perp}<t$. For a given value of $t_{\perp}$, the energy costs of a single local spin flip in the doped and undoped cases can be compared to determine whether they explain the greater hardening observed for small $t_{\perp}$.
In the undoped case, the Heisenberg model gives the following energy cost for a spin flip:
\begin{equation}
\Delta E_{\text{Undoped}}=\frac{4}{U}\left(t^2+t_{\perp}^2\right).
\end{equation}
When one hole is doped on the NN site of the flipped spin, whether the doped hole is on the same chain as the flipped spin or on its neighboring chain must be taken into account.  

While the Heisenberg model only includes terms of order $t^2$ and $t_{\perp}^2$, the three-site term includes three different channels:
\begin{align}
\Delta E_{Doped}=\frac{1}{8}\frac{4}{U}\left(7t_{\perp}^2+4tt_{\perp}+7t^2\right).
\end{align}
Thus, the local picture predicts that for sufficiently small $t_{\perp}$, $\Delta E_{\text{Doped}}<\Delta E_{\text{Undoped}}$, so the spectrum should soften upon doping and harden only for large enough $t_{\perp}$.  Moreover, as $t_{\perp}$ gets closer to the two-dimensional limit, the spectrum should harden more. This is clearly very different from what is observed in Fig.~\ref{fig:xdyn_doping}, where the spin excitations harden less and less upon doping as $t_{\perp}$ increases. 

The local picture may fail because of the high temperature.  As high-energy spin excitations in the low-$t_{\perp}$ regime retain their collective one-dimensional properties, a picture of local magnon creation does not apply. 
In order to examine the prediction of the transition from a softening to a hardening of the spin excitations, a future study could examine the small-$t_{\perp}$ regime at lower temperatures.  If this transition is observed, it will validate the local picture.

\subsection{Comparison with Cluster Perturbation Theory}
\begin{figure*}[t!]
	\includegraphics[scale=0.5]{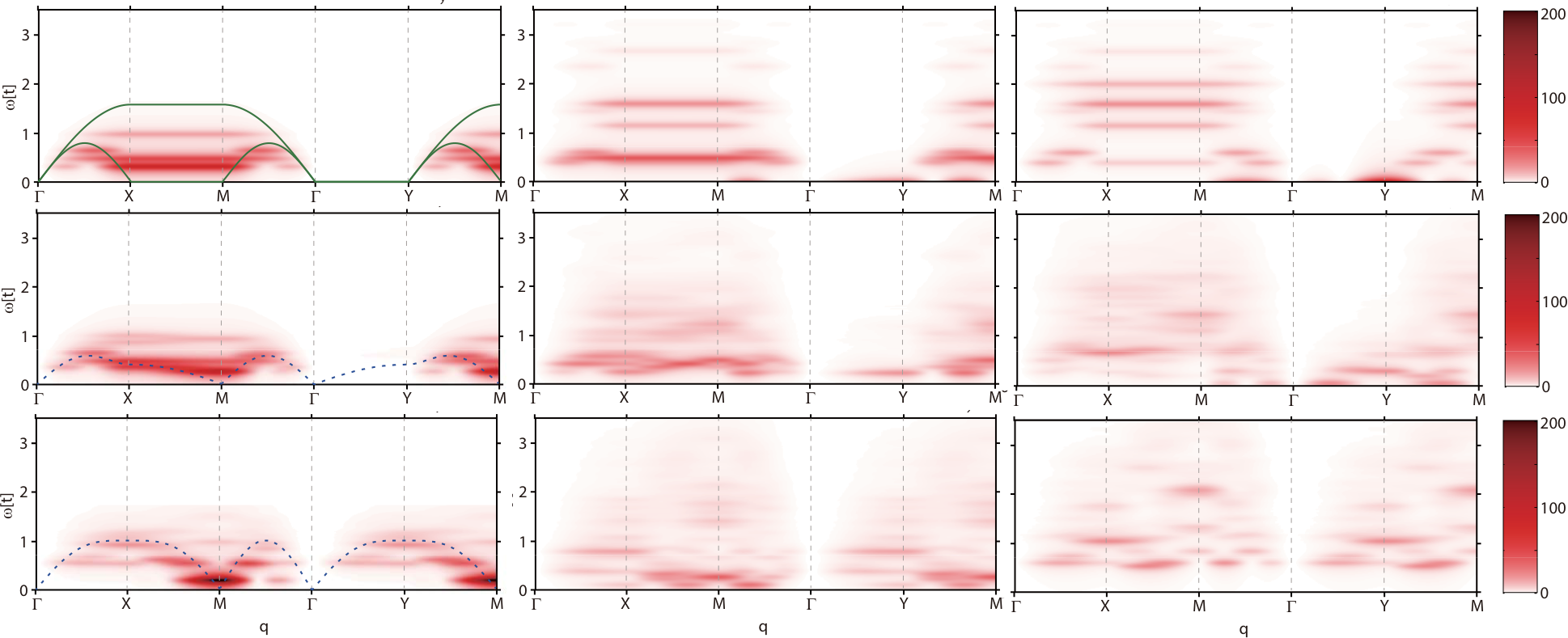}
	\caption{False color plot of the dynamical spin structure factor $S(\textbf{q},\omega)$ along the main symmetry directions, at half-filling (left column), $16.7\%$ (center column), and $33.3\%$ (right column) hole doping, for different values of the transverse hopping integral $t_{\perp}$. The first row corresponds to $t_\perp=0$, the second to $t_\perp=0.4t$, and the third to $t_\perp=t$.  The color scale is the same for all plots.  At half-filling, the solid line corresponds to the two-spinon continuum and the dashed line to the linear spin-wave dispersion.}
	\label{fig:CPT}
\end{figure*}

The DQMC studies of the dimensional crossover in the spin excitation spectrum can be complemented with cluster perturbation theory (CPT) calculations.  This numerical technique combines exact diagonalization (ED) and perturbation theory, dividing the infinite plane into smaller identical clusters that are solved exactly using ED.  Hopping between the clusters is treated to leading order in perturbation theory.  CPT is exact in the limits of strong and weak coupling as the number of Brillouin zone sites $L\rightarrow \infty$.  Unlike DQMC, it is generally performed at zero-temperature, thus avoiding finite-temperature effects.  It also complements DQMC with its fine momentum resolution.  However, because the ED solver works in the canonical ensemble, dopings are limited to discrete levels, as opposed to the continuous doping evolution accessible to DQMC, which works in the grand canonical ensemble.  In this section, the CPT simulations are performed with $U=8t$ and $t_\perp/t=0.0,0.1,0.2,0.3,0.4,0.5,0.6,0.7,0.8,0.9,1.0$.  Both the 12-site C4 and $2\times6$ two-leg ladder systems are used; the results are qualitatively the same.  

Figure~\ref{fig:CPT} shows the dynamical spin structure factor on a C4 cluster with three different doping levels and $t_\perp/t=0,0.4,1.0$ for comparison to Fig.~\ref{fig:xdyn_doping}.  At half-filling, as in the DQMC calculation, the one-dimensional spectral intensity follows the two-spinon continuum (solid lines).  When $t_\perp>0$, the numerical calculation agrees well with the linear spin-wave dispersion (dashed lines) throughout the Brillouin zone, again similar to the behavior of the DQMC calculation.  The only significant discrepancy between the CPT and DQMC spin excitation spectra occurs in the quasi-one-dimensional system at $(0,\pi)$ (the Y-point), where the dynamical spin structure factor from CPT follows the spin-wave dispersion while that from DQMC has spectral intensity at lower energies up to large values of $t_\perp$.  As DQMC is performed at a significantly higher temperature, the difference is most likely due to a thermal effect.

When the system is doped away from half-filling, the CPT and DQMC calculations continue to agree well.  Regardless of doping level, the dimensional crossover occurs in a smooth transition as $t_\perp$ is varied.  When the interchain hopping is increased, the transverse dispersion ($\Gamma \rightarrow Y$) hardens, and spectral weight shifts from $(\pi,0)$ towards $(\pi,\pi)$ as two-dimensional character is enhanced.  In addition, increasing hole doping causes the spectra to harden along the longitudinal direction ($\Gamma \rightarrow X$).  The close agreement of results from these two techniques suggests that they access the same physics despite the difference in simulation temperatures.  

Section~\ref{sec:crossover} has examined the dimensional crossover in half-filled and doped systems using DQMC, complemented by CPT calculations.  The evolution of the doping dependence with increasing interchain coupling has been explored systematically.  The crossover appears to occur smoothly for all doping levels and momentum points.  By comparing the momentum dependence of the crossover in the spin dynamics at half-filling and at $40\%$ doping, the presence of persistent magnon-like excitations has been confirmed. Moreover, the importance of the three-site term in understanding spin dynamics has been highlighted.  The mechanism behind spectral hardening upon hole doping remains unclear, however; future studies could be performed at lower temperatures to study the applicability of the local picture.  The doping trend of the NN spin correlation function as the system evolves from one to two dimensions also sheds light on how doped holes interact with the magnetic order.  Finally, a comparison of spin dynamics calculated by DQMC and CPT forms a more complete picture of the doping-dependent dimensional crossover.

\section{Conclusions}
This study has systematically examined the magnetic properties of the Hubbard model in one and two dimensions and explored the dimensional crossover in the half-filled and doped systems.

In one dimension, the spin dynamics of a strongly-correlated Hubbard chain has been explored for different doping levels.
To understand the role of short-range correlations in the spin dynamics, the doping evolution of the NN spin correlation function has also been examined.  In contrast to the two-dimensional system, doped holes appear to interact weakly with the magnetism at low doping levels, but have an enhanced impact at high doping levels.  A simple picture of spin-charge separation cannot explain the trend for all doping levels. 

In two dimensions, the doping evolution of the spin dynamics has been calculated and compared to an earlier study in which NNN hopping $t'$ was included.\cite{cj}  At the AF wavevector, the spin excitations almost disappear upon doping, as $(\pi,\pi)$ AF order is destroyed.  Along other directions, the spectrum hardens with a persistent intensity. 
The NN spin correlation function reveals that delocalization of doped holes destroys magnetic order in a more subtle way than predicted by a simple local picture of AF exchange.  At lower doping levels, magnetic order is suppressed below what is expected, while at higher doping levels, the local picture appears to apply.

The {\it crossover} of magnetism from one to two dimensions provides a means of elucidating the processes involved in spin dynamics.  
When the dimensionality is tuned between these limits, the magnetic properties drastically change regardless of the doping level.
Crucially, this suggests that the spin excitations in the two-dimensional doped Hubbard model cannot be explained using the spinon picture known from one dimension.
More precisely, we note that:

Firstly, doping modifies the $t_{\perp}$ dependence of the crossover at low, but not high, energies. Indeed, doping enhances the spin dispersion perpendicular to the chains, which can be understood with the three-site term of the $t-J$ model and demonstrates its importance when studying doped systems.\cite{3siteskw}
Comparing the evolution of spin dynamics at half-filling to that at $40\%$ hole doping demonstrates that persistent coherent spin excitations develop at intermediate $t_\perp$ and smoothly interpolate to the two-dimensional case.\cite{cj}  Moreover, the sensitivity of spin excitations to interchain coupling is momentum dependent.

Secondly, dimensionality also affects the way in which persistent spin excitations harden upon doping.  Indeed, the hardening is enhanced with decreasing $t_\perp$.  The local picture used in two dimensions to explain this hardening is adapted to the anisotropic case.  However, for small interchain coupling, it predicts a softening of the spin excitations upon doping, and for large interchain coupling, it predicts that the hardening should increase with $t_{\perp}$.  Evidently, the local picture does not fully account for the hardening mechanism; it may be confounded by thermal effects.  Future studies could simulate the small $t_{\perp}$ part of the crossover at lower temperatures in order to search for this transition from a softening to a hardening.

Finally, the dimensional crossover of the NN spin correlation sheds light on the role of short-range correlations.  Along the chain, it smoothly evolves from the one-dimensional result toward the two-dimensional one as $t_{\perp}$ increases, highlighting the very different nature of the interplay between doping and magnetic order in one and two dimensions.  Perpendicular to the chain, the NN correlation function exhibits a plateau over a large range of doping levels, suggesting that processes may exist to compensate the local reduction of spin density by the doped holes.

In this work, a coupling-driven dimensional crossover approach has been used to calculate the evolution of the spin excitation spectrum for all momentum points.  An alternative strategy would be to study a geometric dimensional crossover tuned by increasing the number of legs in a ladder.\cite{lad5}  This future work would provide an additional perspective on the interplay between dimensionality, doping, and magnetism in the Hubbard model.

This research was supported by the U.S. Department of Energy (DOE), Office of Basic Energy Sciences, Division of Materials Sciences and Engineering, under Contract No. DE-AC02-76SF00515.  Y.F.K. was supported by the Department of Defense (DOD) through the National Defense Science and Engineering Graduate Fellowship (NDSEG) Program and by the National Science Foundation (NSF) Graduate Research Fellowship under Grant No. 1147470.  K.W. acknowledges support by Narodowe Centrum Nauki (NCN, National Science Centre) under Projects No. 2012/04/A/ST3/00331 and No. 2016/22/E/ST3/00560.S.J. acknowledges partial support from The Joint Directed Research and Development (JDRD) program with Oak Ridge National Laboratory.  The computational work was partially performed at the National Energy Research Scientific Computing Center (NERSC), supported by the U.S. DOE under Contract No. DE-AC02-05CH11231.

\bibliography{bibli_bb_KW}

\end{document}